\documentclass[12pt,preprint]{aastex}
\usepackage{natbib}

\begin{document}

\title{Microlensing Parallax for Observers in Heliocentric Motion}

\author{
S.~Calchi~Novati\altaffilmark{1,2,3,a},
G.~Scarpetta\altaffilmark{2,3}
}
\altaffiltext{1}{NASA Exoplanet Science Institute, MS 100-22, 
California Institute of Technology, Pasadena, CA 91125, USA}
\altaffiltext{2}{Dipartimento di Fisica ``E. R. Caianiello'', 
Universit\`a di Salerno, Via Giovanni Paolo II, 84084 Fisciano (SA), Italy}
\altaffiltext{3}{Istituto Internazionale per gli Alti Studi Scientifici (IIASS),
Via G. Pellegrino 19, 84019 Vietri Sul Mare (SA), Italy}
\altaffiltext{a}{Sagan Visiting Fellow}


\begin{abstract}
Motivated by the ongoing \emph{Spitzer} observational campaign, 
and the forecoming K2 one, we revisit, working in an heliocentric 
reference frame, the geometrical foundation for the analysis 
of the microlensing parallax, as measured with the
simultaneous observation of the same microlensing event 
from two observers with relative distance of order AU.
For the case of observers at rest
we discuss the well known fourfold 
microlensing parallax degeneracy and 
determine an equation for the degenerate directions 
of the lens trajectory. For the case of observers in motion, 
we write down an extension of the 
\cite{gould94b} relationship between the 
microlensing parallax and the observable
quantities and, at the same time, we highlight 
the functional dependence of these same quantities
from the timescale of the underlying microlensing event. 
Furthermore, through a series of examples, we
show the importance of taking into account the motion of
the observers to correctly recover 
the parameters of the underlying
microlensing event. In particular
we discuss the cases of the amplitude of the microlensing
parallax and that of the difference of the timescales
between the observed microlensing events,
key to understand the breaking
of the microlensing parallax degeneracy.
Finally, we consider the case of the simultaneous
observation of the same microlensing event from ground
and two satellites, a case relevant 
for the expected joint K2 and \emph{Spitzer}
observational programs in 2016.
\end{abstract}



\section{Introduction}
\label{sec:intro}

The microlensing parallax is a key observable 
to break the degeneracy in the microlensing parameter space 
and recover the physical parameters of the lens,
specifically its mass and distance. The framework 
for the analysis of the microlensing
parallax, through measures from two observers 
distant enough one from the other,
specifically through the simultaneous observation 
of the same microlensing event from ground and from space, 
has been set by \cite{refsdal66} and later put up to
date and developed by \cite{gould94b}.
The era of space-based microlensing-parallax
observations started using \emph{Spitzer} \citep{werner04},
as earlier  suggested by \cite{gould99},
with the analysis of a SMC event \citep{dong07},
and later on continued with the ongoing \emph{Spitzer} 
observational campaign started in 2014
for the follow up of microlensing events
detected towards the Galactic bulge led by A.~Gould
\citep{spitzer14,spitzer15,spitzer16a,spitzer16b}.
This observational campaign  has already led to several important results
assessing clearly the importance of this kind
of measurements, among which
the first microlensing exoplanetary system
with a space-based parallax measurement \citep{ob140124};
the first space-based microlens parallax measurement
of an isolated star \citep{ob140939};
and a first analysis on one of the main goals
of the campaign, the determination
of the Galactic distribution of exoplanets \citep{novati15}
(for this specific issue we also refer
to the recent analysis by \citealt{penny16}).
Furthermore, the space-based microlensing parallax
is expected in the next few years to play an
increasingly relevant role
for the analysis of microlensing
events, in particular for the characterization
of exoplanets, besides \emph{Spitzer}
also with K2 \citep{howell14} and, 
in the longer term, with WFIRST (\cite{spergel15},
for specific analyses on the microlensing
parallax with WFIRST we refer to \citealt{gould13,yee13}).

A key aspect in the analysis of space-based microlensing parallax 
events is the understanding of the underlying degeneracy 
for the parallax determination: a fourfold degeneracy 
of the direction of the parallax ``vector'' and a twofold
degeneracy of the parallax amplitude which is the relevant quantity
for the determination of the physical characteristics of the lens system. 
As further detailed below, this degeneracy arises because 
the observation of the same microlensing event from two observers 
only partly breaks the degeneracy among the directions of the lens-source 
relative motion, a degeneracy which is instead
complete for the observation by a single observer of 
a single lens microlensing event. 
\cite{gould94b} set the framework, which has been followed thereafter, for
the analysis of the microlensing parallax in the case 
of two inertial observers at rest. 
\cite{gould95c} then addressed also the issue of the parallax degeneracy for
the case of two observers in relative motion, 
but still from within the framework established for the case 
of inertial observers at rest. The results of
these works then became the basis for the analysis 
of the  mechanism for breaking  the parallax degeneracy, 
and in particular for two simulations of the microlensing
parallax signal towards the Magellanic Clouds \citep{boutreux96}
and the Galactic bulge \citep{gaudi97}.
Incidentally, we recall that these are two opposite observational
targets as they lie, roughly, at the ecliptic pole
and along the ecliptic plane, respectively,
the second line of sight being
therefore coplanar with typical
satellite orbits, and specifically
for that of \emph{Spitzer} and \emph{Kepler}.

The recently renewed and growing observational importance
of space-based microlens parallax calls
for a better theoretical understanding of its
underlying mechanisms in particular
for the breaking of the degeneracy
for observers in relative motion.
Addressing this issue is the primary goal
of the present work. Specifically, 
as a main result,
we extend the \cite{gould94b} expression
relating the microlens parallax to the
light curve observable
valid for observers at rest to the 
general case of observers in motion.
Considering primarily the line of sight towards
the Galactic bulge, currently the more
relevant from an observational point of view,
we then present a series of microlensing event
test cases to show the impact
of fully including in the analysis
the motion of the observers
comparing in particular with
the outcome of the \cite{gould95c} analysis.
As detailed below, we perform
our analysis from within an heliocentric frame,
therefore from the ideal point of view of
an observer which can be to excellent
approximation considered inertial.
This choice, opposite to the usual
one of a geocentric point view, 
allows us a more clear and transparent discussion  of the problem.

The microlens parallax is a genuine geometrical effect.
The strength of its measure with simultaneous
ground and space-based observations is then
enhanced by its clean signature on the light curves,
resulting usually in rather precise determinations
(the typical relative error for the parallax 
in single-lens systems in \cite{novati15} is about 10\%).
Besides the error on the single
parallax solutions, however, a major source of uncertainty remains that
associated with the discrete degeneracy in the microlensing
parallax determination. In this framework,
to the extent that they may lead to break this
degeneracy, this is where 
combined space-based analyses, as the expected joint
observations with \emph{Spitzer} and K2,
may be expected to be of key importance.
Considering the determination of the physical
lens parameters, in particular the lens mass,
we recall however that the measure of the microlensing parallax alone,
besides the event duration,
is not sufficient to break the degeneracy in the
microlensing parameter space. The overall error
budget for the physical parameters 
is therefore the result of the combined
effect of all the measurable quantities.
The degeneracy may be broken if, beside the microlensing
parallax, the finite source effect,
and therefore the Einstein angular radius, is also known.
This is routinely measured
in multiple-lens, planetary, systems,
but only infrequently in single-lens ones.
Out of the 2015 \emph{Spitzer} campaign \citep{novati15b}, 
\cite{zhu15} report two such (single-lens system) cases, achieving
a relative error for the lens mass of 15\% and 8\%.
The two planetary systems with \emph{Spitzer}-based
microlens parallax, OGLE-2014-BLG-0124 \citep{ob140124} and
OGLE-2015-BLG-0966 \citep{ob150966}, come with a relative error
in the planet mass determination of 30\% and 10\%, respectively where,
in the first case, the error budget is dominated
by the uncertainty related to the finite source effect.
We also recall that an alternative, and for ground-based 
observations only more generally applicable,
channel to the determination
of the physical parameters of lens systems,
independent from the measure of the microlens parallax,
is that associated to the measure,
within a few years of the microlensing event itself,
of both the lens flux and the relative lens-source proper motion 
(the second being key to unambiguously
disentangle the lens and the source flux, \citealt{henderson14}).
This kind of analysis requires high angular resolution imaging facilities,
ground-based adaptive optic systems or space observatories, eg  the \emph{HST}:
in this framework \cite{bennett15} and \cite{batista15} have 
recently reported on their analyses of the planetary 
system OGLE-2005-BLG-169 \citep{gould06} where
they reach a precision of 6\% in the planet mass determination.

The paper is organized as follows. 
In Section~\ref{sec:pie} we 
set the framework of our analysis;
in Section~\ref{sec:rest} we describe
the case of the microlensing parallax
for two observers at rest within an
heliocentric frame, and in particular
we go through a detailed geometrical analysis 
of the underlying degeneracies;
in Section~\ref{sec:motion} we extend
the analysis to the case of observers in motion;
in Section~\ref{sec:ana} we present an
analysis of the parallax for a sample
of light curves, for the line of sight towards
the Galactic Bulge; finally, in Section~\ref{sec:k2}
we discuss the case for the simultaneous
observation of the same microlensing event
from ground and two satellites, which is the
exciting case we expect to happen during the
K2 and \emph{Spitzer} microlensing
campaign in 2016.


\section{The microlensing parallax}
\label{sec:pie}

We start by setting the framework
and the notation of our analysis.
Through the paper we consider a single-point source
single-lens system. A microlensing event
(for a review see for instance \citealt{mao12})
is then characterized by three parameters:
the time at maximum magnification, $t_0$,
the impact parameter, $u_0$, which sets
the magnification at maximum along the
microlensing light curve at $t=t_0$,
and the \emph{Einstein time}, $t_\mathrm{E}$,
which sets the timescale of the event.
The light curve magnification, the typical
bell-like symmetric shape known as 
Paczy\'{n}ski light curve \citep{pacz86}, reads
\begin{equation} \label{eq:pacz}
A\left(t\right) = A\left(u(t)\right) = 
\frac{u^2+2}{u\sqrt{u^2+4}}\,,\quad 
u(t) = \sqrt{u_0^2+\left(\frac{t-t_0}{t_\mathrm{E}}\right)^2}\,,
\end{equation}
where $u(t)$ describes the relative distance,
projected on the lens plane orthogonal
to the line of sight to the source,
of the lens with respect to the observer.

The expression for $u(t)$ in Eq.~\ref{eq:pacz} 
holds for linear uniform lens motion with
respect to an inertial observer; 
it is an excellent approximation for typical
microlensing events, lasting from days up to few months, 
for an ideal observer on the Sun. 
Furthermore, Eq.~\ref{eq:pacz} works well also 
for observers on Earth, except for very long timescale events,
where the orbital motion comes into play.

All the physical parameters of the source-lens
system are enclosed in the event timescale, 
the \emph{Einstein time}
$t_\mathrm{E}=R_\mathrm{E}/v$, where
$v$ is the relative lens-source velocity on the lens plane.
The \emph{Einstein radius}, $R_\mathrm{E}$, 
is the characteristic length of the system,
with all the physical lengths
on the lens plane being normalized with it.
The Einstein radius is a function of the distance from the
observer to the lens, $D_l$,  the source, $D_s$,
and of the lens mass, $M_l$. 
The \emph{microlensing parallax}, $\pi_\mathrm{E}$,
is defined as the inverse of the Einstein radius,
in units of AU, projected on the observer plane,
$\pi_\mathrm{E} = \mathrm{AU} (1-x)/R_\mathrm{E}$, where
$x\equiv D_l/D_s$. We recall that for Galactic
bulge events typically $R_\mathrm{E}$ is of order AU.
(This is the underlying reason
why, to measure the microlensing parallax,
the two observers must lie at about a
relative distance of order AU. The exception are
the rare cases of extremely highly magnified 
microlensing events for which the microlensing parallax 
can be determined from two observers separated by about 1
Earth radius, \citealp{gould97b}).
We recall that the choice of the microlensing parameters
is not univocal, in particular  
the formalism can be more suitably (for several applications)
recasted in terms of observables \citep{gould00b}.

In principle one may further characterize 
a microlensing event by giving, additionally, 
the direction of the lens relative  motion in the lens plane, 
which however remains completely undetermined in Eq.~\ref{eq:pacz}, 
which only contains the modulus of the impact parameter. 
The direction of motion is however relevant when
discussing the parallax, so that we here introduce, 
as a fourth parameter, the angle $\chi$ 
between a fixed direction (in our notation,
the $x$-axis of the reference frame 
to be introduced in Section~\ref{sec:rest}) 
and the orthogonal to the direction of the lens motion.

The three parameters $t_0,\,u_0$ and $\chi$
all characterize the geometry of the microlensing event.
The first two do not carry any information
on the physical parameters of the lens system,
their underlying distribution is indeed flat.
The case of $\chi$ is different, though,
as its distribution reflects that of the underlying
lens and source velocity
and therefore is endowed with an intrinsic,
often relevant, physical information\footnote{From an
observational point of view also the distribution
for $u_0$ is not flat. This reflects both
the efficiency of a given instrumental
setup and therefore, indirectly,
also the underlying source luminosity function.}.

Note finally that in the following we will
always refer to the microlensing parameters
$t_0,\,u_0$ and $t_\mathrm{E}$, without further
subscript, as those of the underlying 
microlensing event as would be seen from the ideal
observer on the Sun.


\subsection{Observers at rest in an heliocentric reference system} 
\label{sec:rest}

In this section we revisit the analysis 
of the microlensing parallax with observers
at rest within an heliocentric framework.
This approach provides us with a different point of view
on well known results and, at the same time,
leads us to highlight some relationships
which, to our knowledge, may not have been already discussed.
Furthermore, this gives us the necessary basis
for the discussion of the case with observers in motion.

We are going to consider an heliocentric cartesian frame
on the lens plane, centered along the line of sight
to the source, as seen from the Sun. The choice of the reference frame
within the lens plane is then arbitrary modulus
a rotation in this plane. The orbital motions
of the observers we are going to discuss all take
place on the ecliptic plane which therefore
takes a privileged position. 
The ``canonical'' choice for the reference frame 
in the lens plane, from a geometrical point of view, 
would then be that of having one of the axes pointing
along the line of nodes, intersection of the ecliptic plane 
with the lens plane. Following the established habit in literature, 
however, we fix the $x$-axis along the line of nodes 
intersection of the lens plane with the equatorial plane instead,
with in particular the $x$ and $y$ axes pointing along 
the equatorial coordinates, west and north, respectively. 
We note however that because the line of sight we consider is
pointing towards the Bulge, roughly on the ecliptic plane 
and at $\lambda \sim 270^\circ$,
these two choices almost coincide.

Let us consider an observer at rest out of the origin, 
which can be the Earth approximated at rest.
Its position in the lens plane, as given by the intersection
of its line of sight to the source
with the lens plane, then depends,
besides from its position in the observer plane,
only on the microlensing parallax.
Specifically, its distance from the origin scales with the microlensing 
parallax\footnote{From an analytical
point of view, this holds within the usual approximation
of neglecting the lengths in the observer
plane, of order AU, compared to the distance
to the source, of order kpc. We note that this
is the same approximation within which we can mix up
geocentric and heliocentric equatorial coordinates.}.
Because of his offset from the origin,
this observer would observe
the same microlensing event as seen
from the (ideal) observer in the origin,
with the same timescale $t_\mathrm{E}$,
but with different impact parameter
and time at maximum magnification.
Let introduce a second observer 
lying into the same observer plane, which
for definiteness we are going to identify
with the satellite \emph{Spizter}, and
which for now we also consider to be at rest.
The relative distance between
the two observers, a known quantity
in the observer plane, when projected in the
lens plane also scales with the microlensing parallax \citep{gould94b}
\begin{eqnarray} \label{eq:gould94}
&&\vec\pi_\mathrm{E} = \vec\pi_\mathrm{E,\pm,\pm} = \frac{\mathrm{AU}}{D_\perp} 
\left(\tau, \Delta u_{0,\pm,\pm}\right)\,,\\
&& \tau = \frac{t_{0,2}-t_{0,1}}{t_\mathrm{E}}\,, \quad 
\Delta u_{0,\pm,\pm} = \pm (|u_{0,2}| \pm |u_{0,1}|)\,. \nonumber
\end{eqnarray}
Here $t_{0,1},\,t_{0,2}$ and $u_{0,1},u_{0,2}$
indicate the time at maximum magnification
and the impact parameter of the
microlensing event as seen by the two observers,
and $D_\perp$ is the relative physical distance between 
the two observers in the lens plane, expressed in AU.
For reference, in the following we will always
identify the observer ``1'' as that on Earth.
This expression is key for the measurement
of the microlensing parallax which is there expressed
in terms of all observable quantities.

\begin{figure}
\epsscale{.90}
\plotone{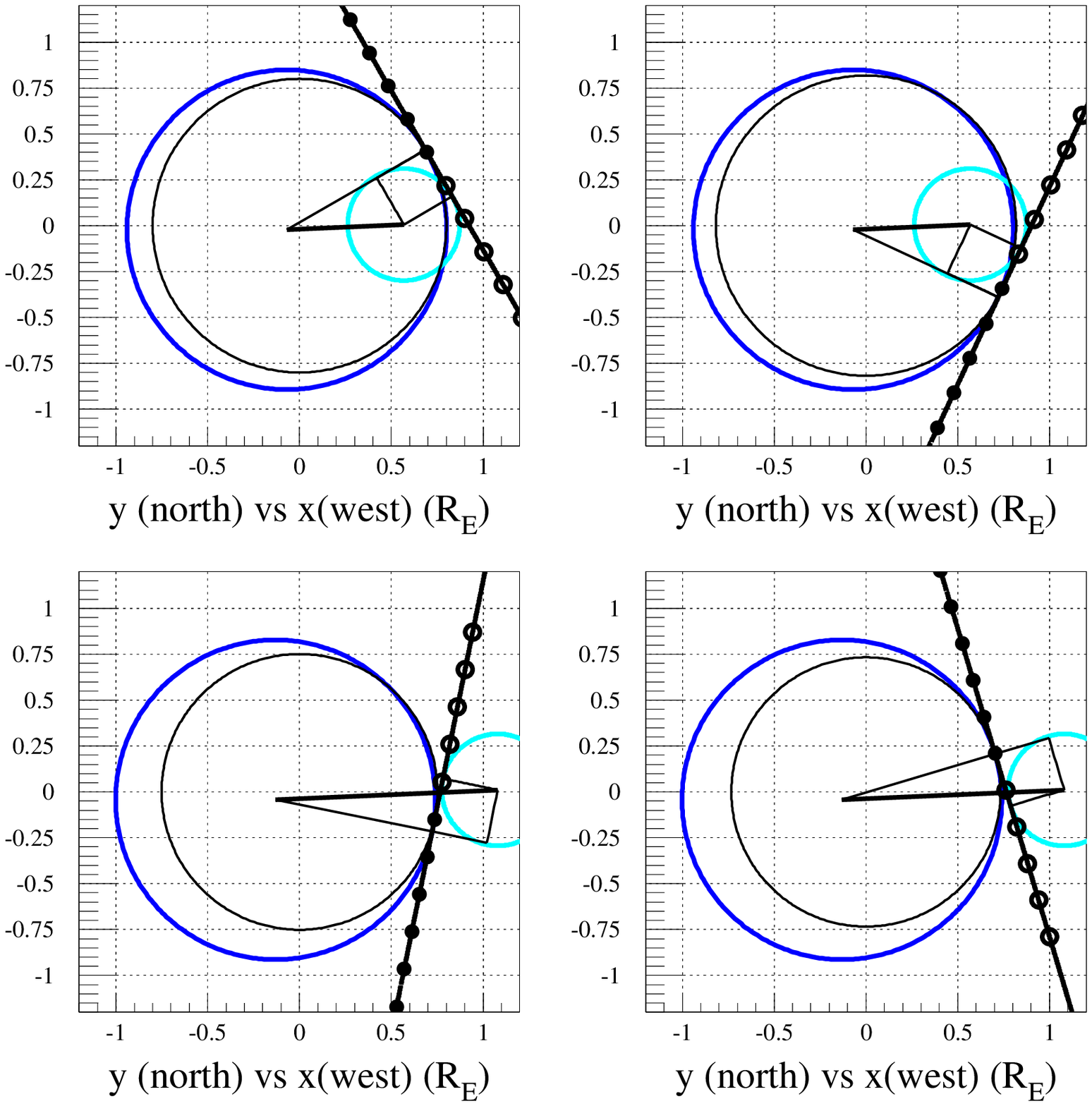}
\caption{Projected in the lens plane,
in an heliocentric reference frame
centered on the line of sight to the source,
the 4 degenerate configurations for the same 
microlensing light curves
as observed by two observers at rest on Earth
and \emph{Spitzer} for a line of sight towards the Bulge
(see  Section~\ref{sec:rest} for full details).
Top (bottom) panels for the 
$\pi_\mathrm{E,-}$ ($\pi_\mathrm{E,+}$) configurations, respectively.
The thin black circle is centered on the origin,
radius the impact parameter as would be seen from the Sun.
The thick circles are centered on the observer positions,
with radius the respective impact parameters,
dark and light blue for Earth and \emph{Spitzer}.
The straight lines simultaneously tangent to the three circles
represent the lens trajectory,
the dots along it are equally spaced by 5 days,
empty ones for times prior $t_0$.
The centers of the Earth and \emph{Spitzer} circles
are joined by a thick line,
whose length scales with the microlensing
parallax; the thinner lines indicates the triangle
construction underlying Eq.~\ref{eq:gould94}.
The $x$ and $y$-axes are along the equatorial 
directions (west and north, respectively), the
$z$ axis along the line of sight, as seen
from the Sun. 
}
\label{fig:prest}
\end{figure}

The basis of Eq.~\ref{eq:gould94} 
lies on elementary geometrical 
considerations which can be done
for instance looking at Figure~\ref{fig:prest},
further discussed below.
The notations $\pm,\pm$ in Eq.~\ref{eq:gould94}
refers to the aforementioned fourfold degeneracy
in the vector parallax determination,
and twofold degeneracy in the modulus, $\pi_\mathrm{E,\pm}$,
which we now describe in some detail.

In Figure~\ref{fig:prest} 
we represent the (four degenerate configurations for the) 
geometry of a microlensing event projected on the lens plane
as seen by the two observers (Earth and \emph{Spitzer}).
Centered on the origin we draw a circle
of radius $u_0$.  From the point of view
of the ideal observer in the origin (the Sun) the observed
light curve is compatible with whatever lens motion
direction tangent to this circle.
The angle $\chi$, though, singles out a unique lens direction,
with the lens passing at the tangent point at time $t_0$. 
We can draw similar circles
around the observer positions on the lens plane, with radius
$u_{0,1}$ and $u_{0,2}$ respectively.
Geometrically, the lens trajectory must then be tangent 
simultaneously to all the three circles,
with $t_{0,1}$ and $t_{0,2}$ being the times
of passage of the lens at the respective
tangent points. 

Following the habit (and with some abuse of notation)
in Eq.~\ref{eq:gould94} 
we have introduced the parallax ``vector'',
$\vec \pi_\mathrm{E}$, with components projected along 
and perpendicular to the lens motion.
The amplitude of the parallax vector, 
the microlensing parallax $\pi_\mathrm{E}$, 
is then obtained applying the Pythagoras theorem to 
the right angle triangle whose hypotenuse
is given by the distance between the two observers, 
and whose cathetus are equal respectively to the distance 
between the tangent points of the two observer circles
and to the difference of the observers impact 
parameters (Figure~\ref{fig:prest}, top panels).

Let us pause to write down an expression
for $\tau=\Delta t_0/t_\mathrm{E}$
as a function of the parameters of the
underlying microlensing event
\begin{equation} \label{eq:tau}
\tau = \pi_\mathrm{E}\,\left(\cos(\chi) \Delta y_0 - 
\sin(\chi) \Delta x_0 \right)\,,
\end{equation}
where $(\Delta x_0,\,\Delta y_0)$ are the distances
of the observers position projected on the lens plane,
with $D_\perp$ in Eq.~\ref{eq:gould94} equal
to $\sqrt{(\Delta x_0)^2+(\Delta y_0)^2}$.
In Eq.~\ref{eq:tau}  $\Delta t_0,\,\Delta x_0$ 
and $\Delta y_0$ are all to be intended 
as signed quantities, as well as $t_\mathrm{E}$, 
whose sign can be thought to identify
the versus of motion along a given direction.
Specifically, our sign convention 
is that, looking the lens plane 
as in Figure~\ref{fig:prest},
the lens motion is anti-clockwise
at the tangent point  between the lens trajectory
and the circle of radius $u_0$ centered in the origin.
We also note that $\tau$ is invariant upon
change of the direction of motion
(which corresponds to a simultaneous change
of the sign of $t_\mathrm{E}$ and therefore also of $\Delta t_0$).
Eq.~\ref{eq:tau} follows  from the geometry of the problem
and is therefore implicit in Eq.~\ref{eq:gould94}
(in particular it is  closely
related to Eq.~8 in \citealt{gould04})
however to our knowledge it was not
previously explicitly written down.
In particular it relates the projections
of the components of $D_\perp$ along
the lens trajectory to the distance between the two tangent points 
to the observer circles, which is equal to the difference 
of the observers time at maximum magnification in units of 
the Einstein time (we recall that all the distances are 
normalized by $R_\mathrm{E}$).
This equation is important as, together with Eq.~\ref{eq:gould94},
it provides the basic analytical understanding
of the underlying fourfold
degeneracy. Specifically, it makes transparent
that $\tau$ only depends, besides
the known observers position, from $\pi_\mathrm{E}$
and $\chi$. Namely, besides being independent
from the event timescale, which
is obvious as the parallax is intrinsically a static quantity,
$\tau$ is also independent from the 
impact parameter and the time at maximum
magnification of the underlying  microlensing event.
This is key to explain the parallax degeneracy.
Fixed $\tau$ and $\pi_\mathrm{E}$,
Eq.~\ref{eq:tau} can be looked at as an equation
for $\chi$, and, in agreement with the geometry
shown in Figure~\ref{fig:prest}, it is a quadratic equation
(see also \cite{gould14} for a discussion of the widespread
appearance of quadratic equations
in microlens parallax). Namely, there are two possible lens trajectories
compatible with the light curves
as seen from the two observers, corresponding
to the two simultaneous tangents to the
observers circles. From a geometrical
point of view the degeneracy follows 
from that we can not establish whether the observers lie
both ``above'' or ``below'' the lens trajectory
(Figure~\ref{fig:prest}, top panels).
Analytically, the key point is the freedom
left by the independence of $\tau$ from $u_0$ and $t_0$
(and $t_\mathrm{E}$). Once fixed $\chi$
according to Eq.~\ref{eq:tau} we can,
independently, fix $u_0$ so that the trajectory is
indeed tangent to the two observers circles,
so to recover the observable values $u_{0,1}$ and $u_{0,2}$,
and, furthermore, suitably shift $t_0$ so to get the same 
observable times at maximum magnification
(for a given value of the timescale).
Finally, we note that following
the habit, we consider the observer
impact parameters as signed quantities,
although, as geometrically they express
a distance, they are intrinsically positive.
The sign indicates whether the observer position
at maximum magnification
lies in the same semi-plane
set by the lens trajectory as the origin, or not
(the sign being then conventionally
positive or negative, respectively),
defining a kind of ``parity'' for the configuration.

The configuration for the two top
panels in Figure~\ref{fig:prest}, 
with the observers lying both in the same semiplane
with respect to the lens trajectory 
(which in principle can be the same
or not as that as the observer in the origin)
is said $\pi_\mathrm{E}=\pi_\mathrm{E,-}$,
with the subscript $-$ to indicate
that in Eq.~\ref{eq:gould94} 
we take the difference of the impact parameters.

Given the same values for the observer impact
parameters and $\tau$,
from Eq.~\ref{eq:gould94} we can then
evaluate a second value for the parallax
amplitude, $\pi_\mathrm{E}=\pi_\mathrm{E,+}$,
taking now the sum of the observers impact parameters.
This case corresponds to the geometry
configurations shown in Figure~\ref{fig:prest}, bottom panels,
with the observers now lying on the  opposite
sides of the lens trajectory.
The new directions for the lens trajectories
can be determined by Eq.~\ref{eq:tau}
for which again we can find suitable
values of $u_0$ and $t_0$ to reproduce exactly
the same observed light curves.
This completes the four-fold parallax
degeneracy for observers at rest.
The geometry for the same couple of $u_{0,1}$ and $u_{0,2}$
is fixed by $\pi_\mathrm{E}$ (two possible values),
$\chi$ and $u_0$ (for possible values, each),
regardless of the event timescale.
Given the geometry, the event timescale 
fixes the observed difference of the times
at maximum magnification.

As a technical point, we note that whereas
Eq.~\ref{eq:tau} can be used only
to determine the directions of the lens
trajectories, the full equations, namely
including $u_0$, can be obtained
from a geometrical analysis looking for
the simultaneous tangents to the
two observers circles
with the constraint that the distance from the two tangent points
must remain equal to $\tau$.
This way, through simple algebra, we can fully analytically
recover the parameters of the 4 degenerate underlying microlensing
events giving rise to the observed light curves.

\begin{figure}
\epsscale{1.0}
\plotone{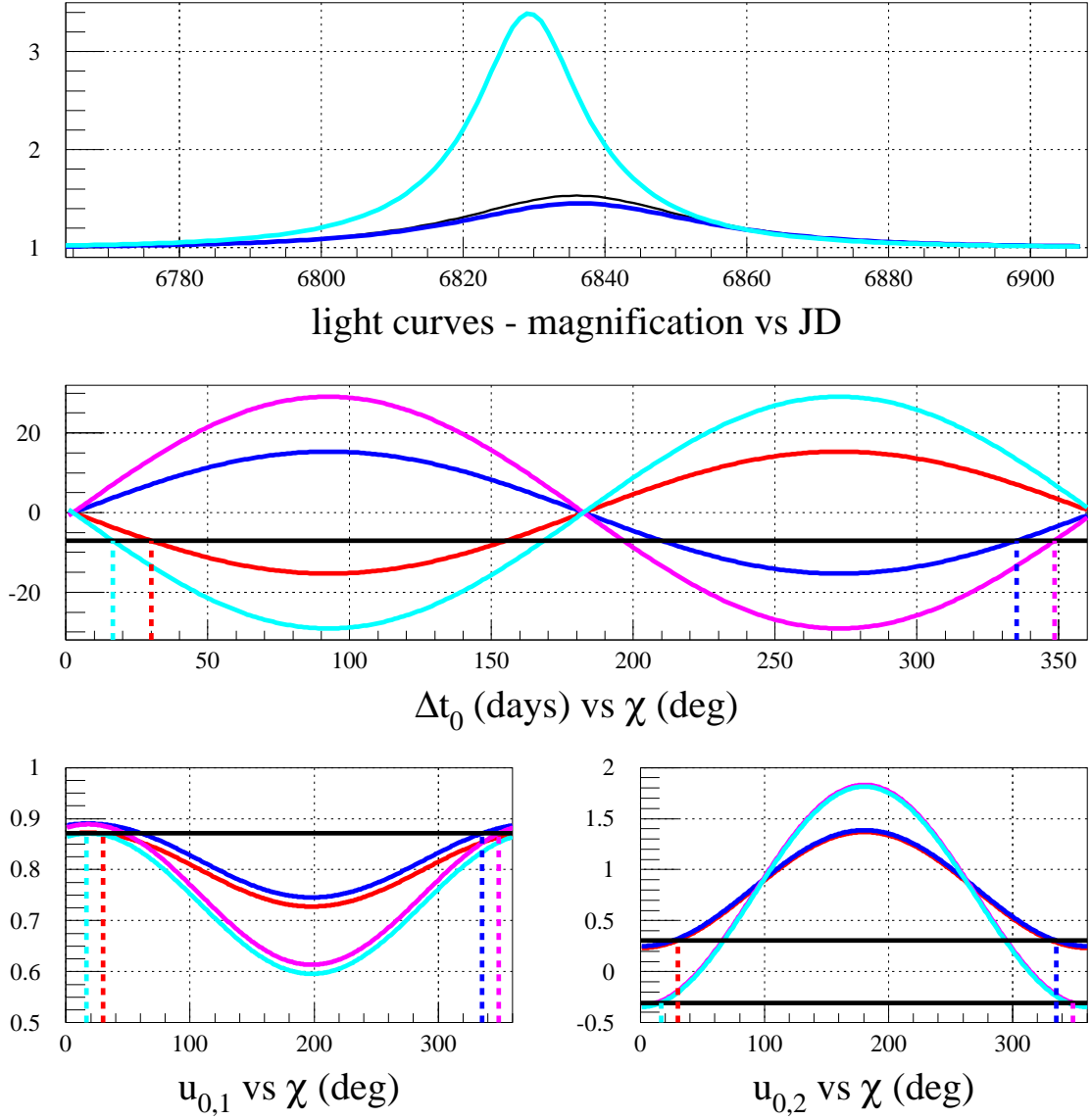}
\caption{Top panel: the light curves corresponding
to the event configurations shown in Figure~\ref{fig:prest}.
The thin black line for one out of four
degenerate solutions as would be seen from the Sun,
in dark and light blue the light curves as seen
from Earth and from \emph{Spitzer}.
Middle and bottom panels:  difference
of the observers times at maximum magnification,
$\Delta t_0$ (middle panel), and 
observer impact parameters, $u_{0(1,2)}$,
for the four degenerate underlying solutions
as would be seen from the Sun,
as a function of the angle identifying the direction
of the lens trajectory, $\chi$.
The solid horizontal lines indicate
the values, for $\Delta t_0$ and $u_{0(1,2)}$, corresponding
to the case shown in Figure~\ref{fig:prest},
with the vertical dotted lines marking
the values of the corresponding angles $\chi$.
See Section~\ref{sec:rest} for full details
on the event configuration.
}
\label{fig:prest2}
\end{figure}

In Figure~\ref{fig:prest2}, top panel,
we show the light curves corresponding to
the four degenerate configurations shown in Figure~\ref{fig:prest}.
In the middle and bottom panels
we fix the configuration to the four
underlying degenerate microlensing events
(therefore for appropriate 
different values for $u_0$ and $t_0$)
and let the angle $\chi$ vary. 
For each given configuration we show the variation for
$\Delta t_0$, middle panel (Eq.~\ref{eq:tau}),
and, bottom panels, for the observers impact parameters,
$u_{0,1}$ and $u_{0,2}$. In particular
$u_{0,2}$ moves also to negative values,
whereas $u_{0,1}$ remains at the same time positive,
corresponding to the configuration
with the two observers lying 
on opposite side with respect to the lens trajectory
(which is indeed always the case for the $\pi_{\mathrm{E},+}$ configurations). 
The dotted vertical lines indicates the values of the angle
$\chi$ corresponding to the degenerate configurations.
We note finally that for each configuration
there are two values of $\chi$ for which we get
the same value for $\Delta t_0$, but these come
with different values of the impact parameters

For reference, the numerical values that
fix the configurations shown in Figures~\ref{fig:prest} and \ref{fig:prest2}
are as follows. The line of sight, towards the Bulge, is fixed
at $\mathrm{RA, DEC} = 266^\circ.8,\,-21^\circ.4$
(with ecliptic latitude $\beta = 2.0^\circ$),
with the source, as hereafter we are
always going to assume, in the Bulge at $D_s=8.~\mathrm{kpc}$. 
We fix (arbitrarily) the observer positions
at the Earth and \emph{Spitzer}
positions along their orbits (discussed in Section~\ref{sec:motion})
at $t=(\mathrm{JD}-2450000) = 6836.0$ (June 27, 2014). 
The four underlying degenerate
microlensing event configurations,
for timescale $t_\mathrm{E}=24~\mathrm{d}$
have parameters (top to bottom, left to right in Figure~\ref{fig:prest})
$t_0 = (\mathrm{JD}-2450000) = 6836.0, \, 6835.17,\, 6834.73,\, 6836.45$,
$u_0 = 0.80,\, 0.82,\, 0.75,\, 0.73$,
and $\chi = 30.0^\circ,\,-24.9^\circ,\,-11.5^\circ,\,16.6^\circ$;
the two parallax amplitude values are $\pi_\mathrm{E,-}=0.60$
and $\pi_\mathrm{E,+}=1.14$. The corresponding observers parameters are
$u_{0,(1,2)}=0.87$ and $0.30$ with
$t_{0,(1,2)}=6836.36$ and $6829.30$ resulting in $\Delta t_0=-7.1~\mathrm{d}$.
The ``sign'' of $u_{0,1}$ is always positive,
that of $u_{0,2}$ is negative for two $\pi_\mathrm{E}$ configurations
(bottom panels in Figure~\ref{fig:prest}).

Finally, note that in Figure~\ref{fig:prest}
we also indicate the direction of motion.
With $\Delta t_0<0$ it results,
according to our sign convention, $t_\mathrm{E}>0$ 
in the top left and the bottom right panels,
$t_\mathrm{E}<0$ in the others.


\subsection{Observers in relative motion} \label{sec:motion}

We now consider the situation for observers in motion. 
The case of a single observer, specifically the effect 
of the Earth orbital motion leading to a deviation from the 
Paczy\'{n}ski shape and its relationship with the microlensing
parallax is known \citep{gould92a}. 
The first measure of a microlensing parallax due
to this effect was reported by the MACHO collaboration \citep{macho95} 
to which many additional cases followed whose physical 
interpretation is not, however, always 
straightforward \citep{poindexter05}.
Moreover (and we recall that we are only discussing 
single-lens systems), the orbital parallax effect
becomes observationally relevant only for a minority of 
unusually long timescale events. 
For theoretical analyses of this effect we also refer to 
\cite{dominik98,smp03,gould04}. Here our goal is to develop 
the analysis for two observers in motion
within the same framework established in Section~\ref{sec:rest}.  
In particular this will lead us to write a 
generalization of Eq.~\ref{eq:gould94}, and Eq.~\ref{eq:tau},
relevant for this case.

Within the heliocentric frame the lens trajectory 
is always a straight line. The motion of the observer 
is taken into account by projecting onto the lens plane
the temporal evolution of his (known) orbital motion 
in the ecliptic plane. Fixed the line of sight, 
this projection then only depends on the microlensing parallax, 
with a circular orbit being generically projected into an ellipse. 
In particular the microlensing parallax 
is equal to the semimajor axis of the
ellipse obtained by projection 
of the observer orbit on the lens plane.

The microlensing light curve is determined by the temporal 
evolution of the lens-observer relative projected distance, 
$u(t)$, according to the same expression of the magnification 
$A(t)=A(u(t))$ as in Eq.~\ref{eq:pacz}
where we have now, however, to take into account of 
the temporal evolution of the observer
position. The combined effect of the two clocks 
in the system, the unknown microlensing timescale which fixes 
the velocity of the lens and the known observer orbital motion, 
makes that the resulting light curve will 
no longer be symmetric around the time at maximum magnification. 
It is still useful, however, to analyse the configuration 
in terms of the circle centered on the observer position 
at the time at maximum magnification with radius given 
by the impact parameter. The specific characteristic for an
observer in motion is that this circle is no longer tangent 
to the lens trajectory, rather, secant. 
Namely, the key point marking out the difference
with respect to the case of an observer at rest is that 
the impact parameter in general does not coincide with 
the minimum geometrical distance from the observer 
projected orbital position at the time at maximum magnification 
to the lens trajectory.

Let us consider two observers in motion.
According to the values 
of the underlying microlensing event,
and specifically  of the microlensing parallax,
we can still find any of the configurations
considered in Section~\ref{sec:rest}
and shown in Figure~\ref{fig:prest}  as for the relative
position of the observers at the time
at maximum magnification with respect to the lens trajectory.
Additionally, the 
relative position of the observers still scales with
the microlensing parallax.
Accordingly, we can write down
an expression similar to Eq.~\ref{eq:gould94}
where we have now, however, to take into account
the effect of the observers motion 
\begin{eqnarray} \label{eq:prof15}
&&\vec\pi_{\mathrm{E},\pm} = \frac{\mathrm{AU}}{D_\perp}\,
\left(\tau - \Delta u_{0,\parallel},
|u_{0,2,\perp}| \pm |u_{0,1,\perp}|\right)\,,\\
&& \Delta u_{0,\parallel} = u_{0,\mathrm{2},\parallel}-u_{0,\mathrm{1},\parallel}\nonumber\\
&& u_{0,\mathrm{oss},\parallel} = |u_{0,\mathrm{oss}}| \sin(\gamma_\mathrm{oss})\,,\quad
u_{0,\mathrm{oss},\perp}=|u_{0,\mathrm{oss}}| \cos(\gamma_\mathrm{oss})\,,
\quad \mathrm{oss}=1,2\,.\nonumber
\end{eqnarray}

We have introduced here the angle $\gamma_\mathrm{oss}$ 
centered on the observer position at the time
at maximum magnification,
between the orthogonal to the lens trajectory 
and the line to the intersection of the lens trajectory
with the circle of radius $u_{0,\mathrm{oss}}$,
so that $u_{0,\mathrm{oss},\parallel}$ and $u_{0,\mathrm{oss}\perp}$
are the projections of the observer impact 
parameter along and orthogonal to the lens trajectory.
In our sign convention, 
for $t_\mathrm{E}>0$ according to the discussion
in Section~\ref{sec:rest}, $\gamma_\mathrm{oss}$  takes
negative (positive) values
when $t_{0,\mathrm{oss}}$, the crossing
time of the lens trajectory with the observer circle,
comes before (after) the time 
at the intersection between the lens trajectory
and the orthogonal from the circle center.
The sign of $\gamma_\mathrm{oss}$  is then reversed for $t_\mathrm{E}<0$.
This way $\gamma_\mathrm{oss}$, independently
of the direction of the lens motion, therefore
of the sign of $t_\mathrm{E}$, always increases anti-clockwise
when looking at the lens plane.

Eq.~\ref{eq:prof15} reduces to that
for the case of observers at rest for $\gamma_\mathrm{oss}=0$,
namely when the lens trajectory
is tangent, and not secant as in this case,
to the circle centered on the
observer position with radius $u_{0,\mathrm{oss}}$.
Note that the $\gamma_\mathrm{oss}$  
are not additional free parameters of the problem,
rather, they are determined by the interplay
within the lens and the orbital motions.
As for Eq.~\ref{eq:gould94}, the components
of the parallax vector are meant to be written in a frame
with the $x$-axis parallel to the lens trajectory.
Specifically, as for the case of observers at rest,
the microlensing parallax is obtained
by application of the Pythagoras theorem.
The difference comes because 
$\tau$ is (by definition) still equal to the (signed) distance
along the lens trajectory between the points
of intersection with the observers circles 
at the time at maximum magnification,
so that the $u_{0,\mathrm{oss},\parallel}$ terms are  
needed to complete the cathetus
delimited by the intersections
to the orthogonals from the
observer positions to the lens trajectory
(in modulus, they can be added or subtracted
depending on the sign of $\gamma_\mathrm{oss}$).

A key point to be stressed for Eq.~\ref{eq:prof15}
is that the notation $\pm$ for $\pi_{\mathrm{E},\pm}$
is only meant to describe the two configurations
with observers on the same ($-$) or
on opposite ($+$)  sides of the lens trajectory.
Because of the observers motion 
these are however no longer degenerate configurations.

Interestingly, in agreement with Eq.~\ref{eq:prof15},
it is possible to write 
an equation relating the relevant lengths
along the lens trajectory analogous of Eq.~\ref{eq:tau}
\begin{equation} \label{eq:tau2}
\tau - \Delta u_{0,\parallel} = \pi_\mathrm{E}\,\left(\cos(\chi) \Delta y_0 - 
\sin(\chi) \Delta x_0 \right)\,.
\end{equation}
Comparing to the case for observers at rest,
on the left-hand side the difference,
the new term $\Delta u_{0,\parallel}$,
follows from the appearance of the angles $\gamma_\mathrm{oss}$;
a key difference is however also found in
the right-hand side, although formally 
identical. Indeed now $\Delta x_0,\,\Delta y_0$,
the relative observer positions
at the time at maximum magnification,
are no longer constant (once given the orbit of the observer),
rather they depend from all the parameters
of the underlying microlensing event configuration,
in particular also from $\chi$, and
indeed the same holds also for the term $\tau$.
This equation therefore no longer identifies
degenerate configurations.
The underlying motivation is that, as discussed, the observed
light curves, fixed the geometrical configuration,
depend from the relative lens-observer motion.
Any variation in the microlensing parameters
is then reflected, in particular, in a change of the observer
positions at the time at maximum magnification
and ultimately in the light curve observables.
In principle,  one can invert this line of thought and claim that there 
is a relationship between the observables, impact parameter 
and time at maximum magnification, and in particular 
the timescale of the underlying microlensing event.


\section{Analysis} \label{sec:ana}

In the previous sections we have revisited
the underlying geometrical and mathematical foundations
for the description of the measure of
the microlens parallax from two distant observers.
In particular we have put in evidence
the differences which arises moving from the case
of observers at rest to that of observers
in relative motion writing down
an extension of the \cite{gould94b}
parallax equation valid for observers at rest
to this case. In this Section first we show
an example of the configuration
described in Section~\ref{sec:motion}
then we highlight,
through specific examples,
the importance of fully taking into account
the relative motion of the observers
for a correct understanding of the underlying
microlensing parallax signal.

As above, for definiteness we consider the case of the simultaneous
observation of the same microlensing event from
ground and from \emph{Spitzer}. \emph{Spitzer} \citep{werner04} moves along
an heliocentric, ``earth-trailing'', orbit, currently
at a distance of about 1~AU from 
Earth\footnote{The ephemeris  of \emph{Spitzer}
as a function of time can be found on the
NASA-JPL Horizon system http://ssd.jpl.nasa.gov/?horizons.}.
For simplicity of the discussion but still accurately enough,
we approximate both Earth's and \emph{Spitzer}'s
orbits as circular, with radius 1~AU and
constant angular velocity with a period of 1~year
and $373~\mathrm{d}$, respectively.
We fix the  Earth and \emph{Spitzer} phases,
their relative azimuthal angles,
at the time of the autumnal equinox.
For 2014, at $\mathrm{JD-2450000}=6923.6$,
$\Delta\lambda = -79.7^\circ$.

\begin{figure}
\epsscale{.90}
\plotone{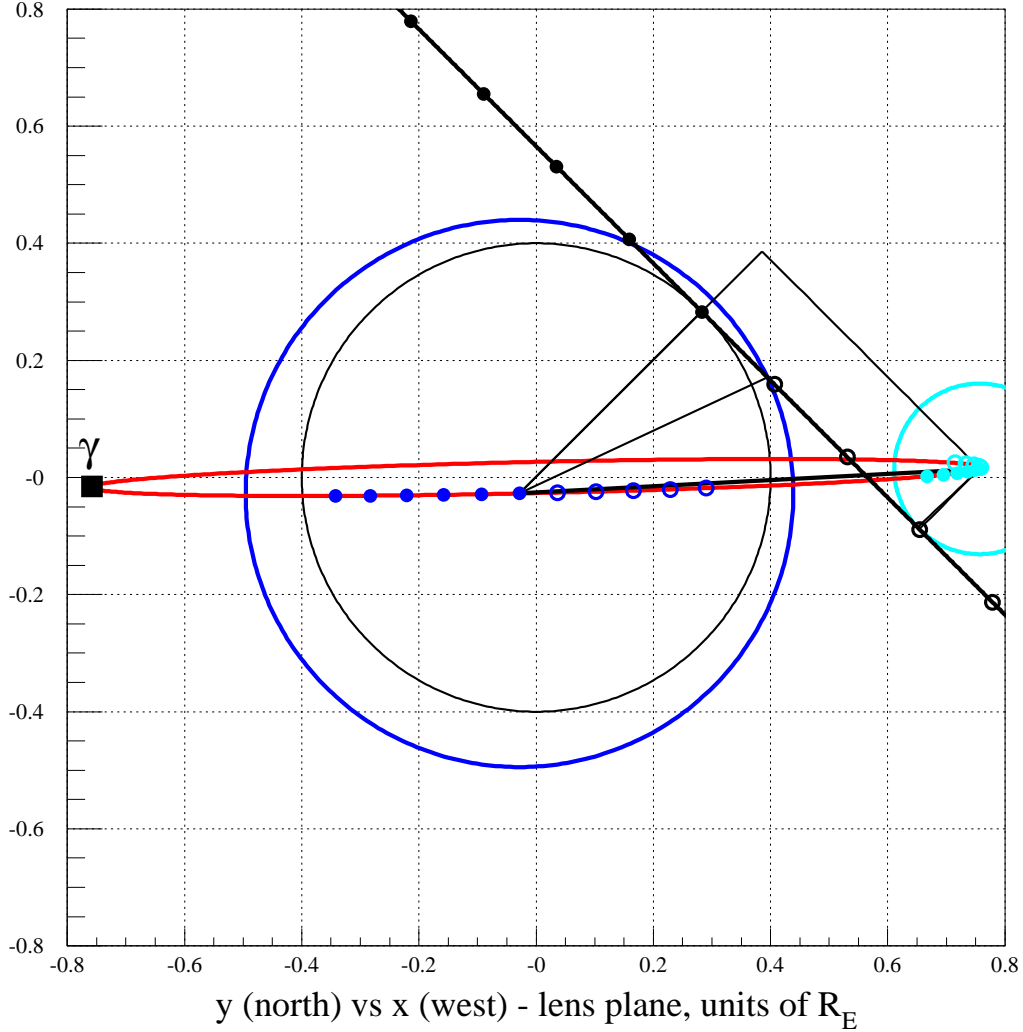}
\caption{Projected in the lens plane,
in an heliocentric reference frame
centered on the line of sight to the source,
the configuration for a parallax
microlensing event for two observers
in motion, on Earth and \emph{Spitzer},
for a line of sight towards the Bulge
(see Section~\ref{sec:motion} for full details). 
The thin black circle is centered on the origin,
radius the impact parameter as would be seen from the Sun.
The thick circles are centered on the observer positions
at the times of their respective maximum magnification
with radius their respective impact parameters,
dark and light blue for Earth and \emph{Spitzer}.
The red elongated ellipse represents the projected 
orbit of the observers
(the square indicates the position
of the Earth at the fall equinox).
The thick straight line
tangent to the circle of radius $u_0$,
secant to the observer circles,
is the lens trajectory. 
The dots along the trajectories are equally spaced by 5 days,
empty ones for times prior the respective
times at maximum magnification.
The centers of the Earth and \emph{Spitzer} circles
are joined by a thick line,
whose length scales with the microlensing
parallax, the thinner lines indicates the triangle
construction underlying Eq.~\ref{eq:prof15}.
The reference frame is as in Figure~\ref{fig:prest}.
}
\label{fig:pmot}
\end{figure}

In Figure~\ref{fig:pmot} we show a specific 
two-observers microlensing parallax configuration
on the lens plane. 
The analysis is carried out following the
discussion in Section~\ref{sec:motion}.
In particular, we evaluate the observers impact parameters
and time at maximum magnifications
through a numerical minimization of
the function $u(t)$
taking into account both the lens and the observers orbital motion.
The ellipse indicates the projection of the Earth,
and \emph{Spitzer}, orbit on the lens plane.
The dotted points along the trajectories
are equally spaced by $5~\mathrm{d}$, with
empty symbols for times prior $t_0$,
showing the direction of motion.
We remark, as discussed in Section~\ref{sec:motion},
that the lens trajectory is \emph{secant}
to the two circles of radius $u_{0,\mathrm{oss}}$
centered on the orbital positions  at time $t_{0,\mathrm{oss}}$
(and no longer tangent as for the case of observers at rest, 
Figure~\ref{fig:prest}).
For reference, the microlensing event configuration
is as follows. The line of sight, towards the Bulge, is fixed
at $\mathrm{RA, DEC} = 266^\circ.8,\,-21^\circ.4$
(we see here a generic feature of the \emph{Spitzer}
observational campaign in 2014 and 2015, with
Earth and \emph{Spitzer} almost aligned
along the equatorial axis).
The microlensing event parameters are 
$t_0 = (\mathrm{JD}-2450000) = 6836.0$, $u_0=0.4$, 
$t_\mathrm{E}=28.5~\mathrm{d}$, 
$\chi=45^\circ$ and $\pi_\mathrm{E}=0.76$ 
which results into a $\pi_\mathrm{E,+}$ configuration,
observers on different sides of the lens trajectory
with $u_{0,(1,2)}=0.47,\, 0.15$
and $t_{0,(1,2)}=6831.5,\, 6821.2$ with $\Delta t_0=-10.3~\mathrm{d}$.
Comparing to the case of observers at rest there
is now, in agreement with Eq.~\ref{eq:prof15},
a non-zero value for $\gamma_\mathrm{oss}$ with
$\gamma_{1,2}=-20^\circ$ and $2^\circ$
(note in particular the negative sign of $\gamma_1$).
As further detailed below, it is useful
to estimate a proxy for the event timescale
analog to the Einstein time for the
observed light curves. In this case
it results $t_{e,oss}$ equal to 36 and 29 days, respectively.

\begin{figure}
\epsscale{1.0}
\plotone{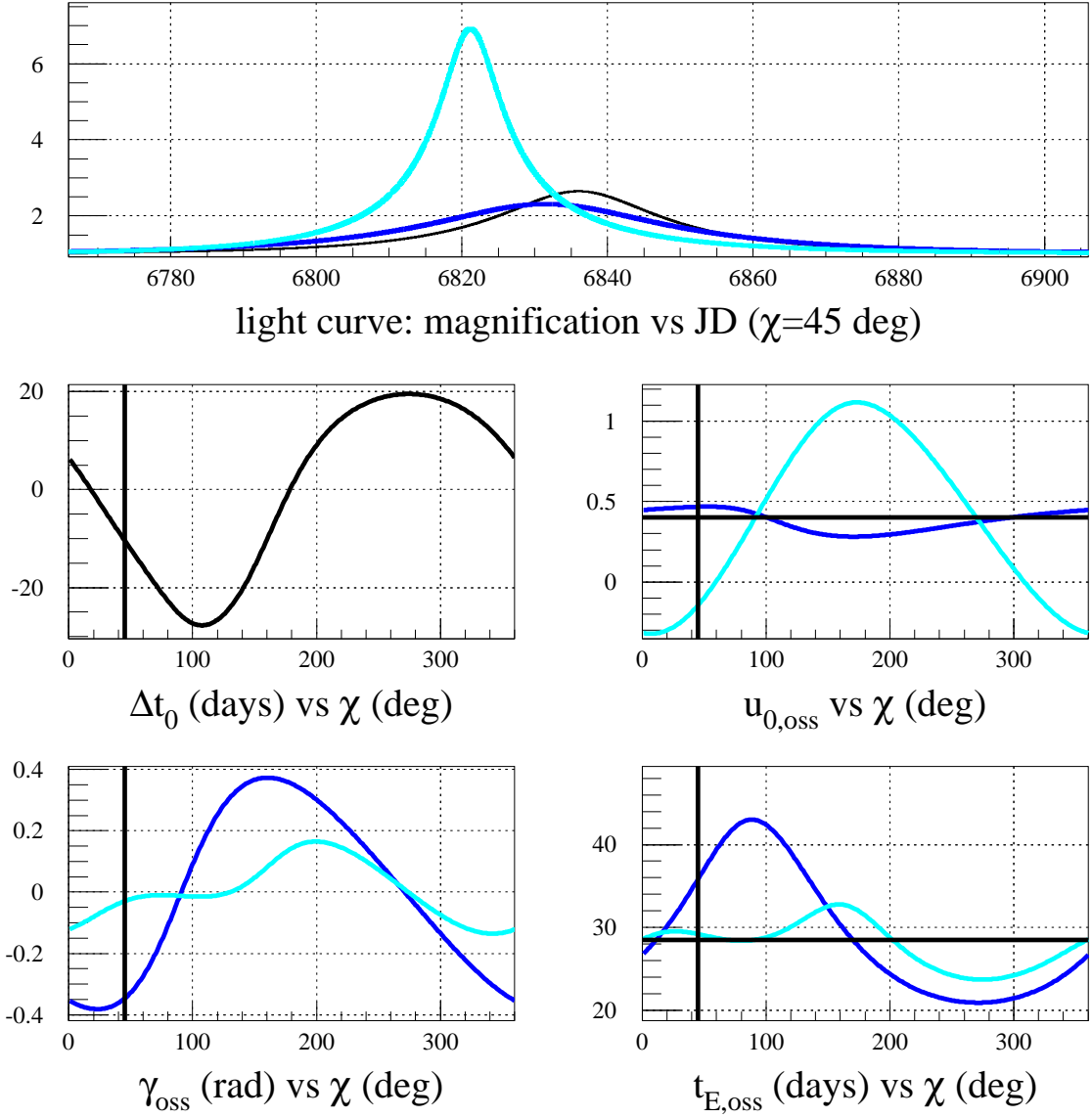}
\caption{Top panel:  the light curves for
the microlensing event configuration shown in Figure~\ref{fig:pmot}.
The thin black line for the event 
as would be seen from the Sun;
in dark and light blue the light curves as seen
from Earth and from \emph{Spitzer}.
Middle and bottom panels: for the same event configuration,
varying the angle $\chi$ of the direction of
the lens trajectory,
difference of time at maximum
magnification, $\Delta t_0$, observers impact parameters, $u_{0,\mathrm{oss}}$,
angle $\gamma_\mathrm{oss}$ and duration 
for the observers light curves, $t_\mathrm{E,oss}$.
The vertical lines indicate the value for $\chi$
used for the light curves in the top panel,
the horizontal lines the values $u_0$ and $t_\mathrm{E}$
of the underlying microlensing event as seen from the Sun.
}
\label{fig:pmot2}
\end{figure}

In Figure~\ref{fig:pmot2} we show, top panel,
the light curves for the microlensing event
shown in Figure~\ref{fig:pmot}  as would be
seen from the Sun, and those for the Earth and \emph{Spitzer} observers.
In the middle and bottom panels we show
a few characteristics quantities
for this microlensing event configuration
by varying the angle $\chi$. In particular,
we show $\Delta t_0$ and $u_{0,\mathrm{oss}}$, no
longer symmetric as in the case with observers at rest
(Figure~\ref{fig:prest}). Furthermore,
we show the characteristics quantities
for observers in motion: the angle $\gamma_\mathrm{oss}$,
which we see can become rather large,
and the estimated timescale along the
observers light curves, $t_\mathrm{E,oss}$.
Note that, besides $u_{0,\mathrm{oss}}$,
also $\gamma_\mathrm{oss}$ is shown
taking into account the sign of the configuration
as defined in Section~\ref{sec:rest}.
The inspection of the bottom-right panel
reveals that the $t_\mathrm{E,oss}$ values 
can get to be significantly different
one from the other, and from the duration
of the underlying microlensing event, $t_\mathrm{E}$.
Indeed, also comparing to Figure~\ref{fig:pmot},
we see that the larger differences for $t_\mathrm{E,1}$
from $t_\mathrm{E}$ occurs for a lens direction of motion
about along the $x$ axis.
This is so because of the Earth position along its orbit
at the time at maximum magnification,
with the relative lens-observer velocity 
getting to a minimum (maximum), and correspondingly
the duration a maximum (minimum) for values about $\chi=\pi/2$ ($3 \pi/2$)
respectively\footnote{Our sign convention 
for $t_\mathrm{E}$, as also apparent from
Figure~\ref{fig:pmot}, is the same as in Section~\ref{sec:rest}.}.

In Section~\ref{sec:motion} we have
discussed the underlying reason
why the microlensing parallax degeneracy 
is broken once introduced the observer motion.
One may however assume that the deviations
from the case with observers at rest 
are in some sense ``small'' and 
look also in this case for the analog of the degenerate configurations
discussed with observers at rest.
More specifically, the ratio of the analysis is the following.
For a given set of parameters we evaluate,
following the analysis in Section~\ref{sec:motion},
the microlensing light curves as seen by the observers in motion,
and in particular the values for 
the impact parameter and time at maximum magnification,
$(u_0,\,t_0)_\mathrm{oss}$. We then consider as given
these values, fix the positions of the observers
along their orbits at their respective time
at maximum magnification and, following the analysis
in Section~\ref{sec:rest}, study the event
under the assumption of observers at rest,
namely through Eq.~\ref{eq:gould94} and Eq.~\ref{eq:tau}.
In particular we can evaluate and compare
the resulting values for the degenerate solutions
of the microlensing parallax 
to the ``true'' value.

Additionally, we consider the observed timescale,
key to address the issue of the breaking of the
parallax degeneracy. Indeed, \cite{gould95c}
acknowledged that the microlensing parallax
fourfold degeneracy is broken as soon as
one drops the assumption of observers at rest
and in particular remarked that the two observers (from ground
and from the satellite) would measure
a different timescale. \cite{gould95c} then
derived a relationship (his Eq.~2.3)
for evaluating the difference
of the observed parameters given
the relative motion of the observers, 
and all in principle directly observables quantities:
the timescale difference
(rather, $\Delta\omega$, 
the difference of the inverse 
of the timescale, $\omega = 1/t_\mathrm{E}$),
and the difference of the times
at maximum magnification. 
This same equation was then the basis 
to address the issue of the possibility 
of breaking the degeneracy 
for the analyses by \cite{boutreux96} and \cite{gaudi97}
with the simulation of parallax microlens events
towards the Magellanic Clouds and the Galactic bulge, respectively.

From the standpoint of our analysis
in Section~\ref{sec:motion}, considering in particular
Eq.~\ref{eq:prof15}, it is however relevant to observe
that the analysis in \cite{gould95c} 
is still based on Eq.~\ref{eq:gould94} valid
for the case of observers at rest.
Looking at the relationship obtained
by \cite{gould95c} as an equation for
$\Delta\omega$, we may therefore
compare this ``expected'' value
to the ``true'' one which we can estimate
from the analysis of the light curves.
This way we can test whether the estimate 
is reliable for assessing the breaking
of the degeneracy.

Before moving on presenting the
results of this analysis we pause
to specify what we intend by ``timescale''
for the case of observers in motion.
The Einstein time, $t_\mathrm{E}$,
as it appears in Eq.~\ref{eq:pacz},
is well defined for a light curve
symmetric around the time at maximum
magnification, $t_0$, and in particular
it results that the magnification
$A(t)$ for $t=t_\mathrm{E}$ 
is that evaluated for
a value $u(t)=\sqrt{u_0^2+1}$.
As detailed above, the light curve
for observers in motion is no longer
symmetric and by itself a single parameter
as the timescale, however defined, can not grasp 
both the width of the light curve
and the degree of asymmetry. Still, it 
remains a useful indicator
of the light curve shape.
Based on the definition
valid for an observer at rest,
as a proxy for the timescale for observers
in motion we proceed as follows.
Given the observable $u_{0,\mathrm{oss}}$
we evaluate the value of the magnification
of an hypothetical Paczy\'{n}ski light curve
with this value for the impact parameter
at the time corresponding to that
of the Einstein time, which we call $\bar{A}$. Moving back
to the observed light curve,
we numerically evaluate the time
interval, in general asymmetric
around the time at maximum magnification,
fixed by the intersection of the light curve
with the value $\bar{A}$ and define
the ``Einstein time'' for the observer in motion, $t_\mathrm{E,oss}$,
as half this interval (with the true $t_\mathrm{E}$
being therefore recovered for a symmetric light curve).

\begin{figure}
\epsscale{1.0}
\plotone{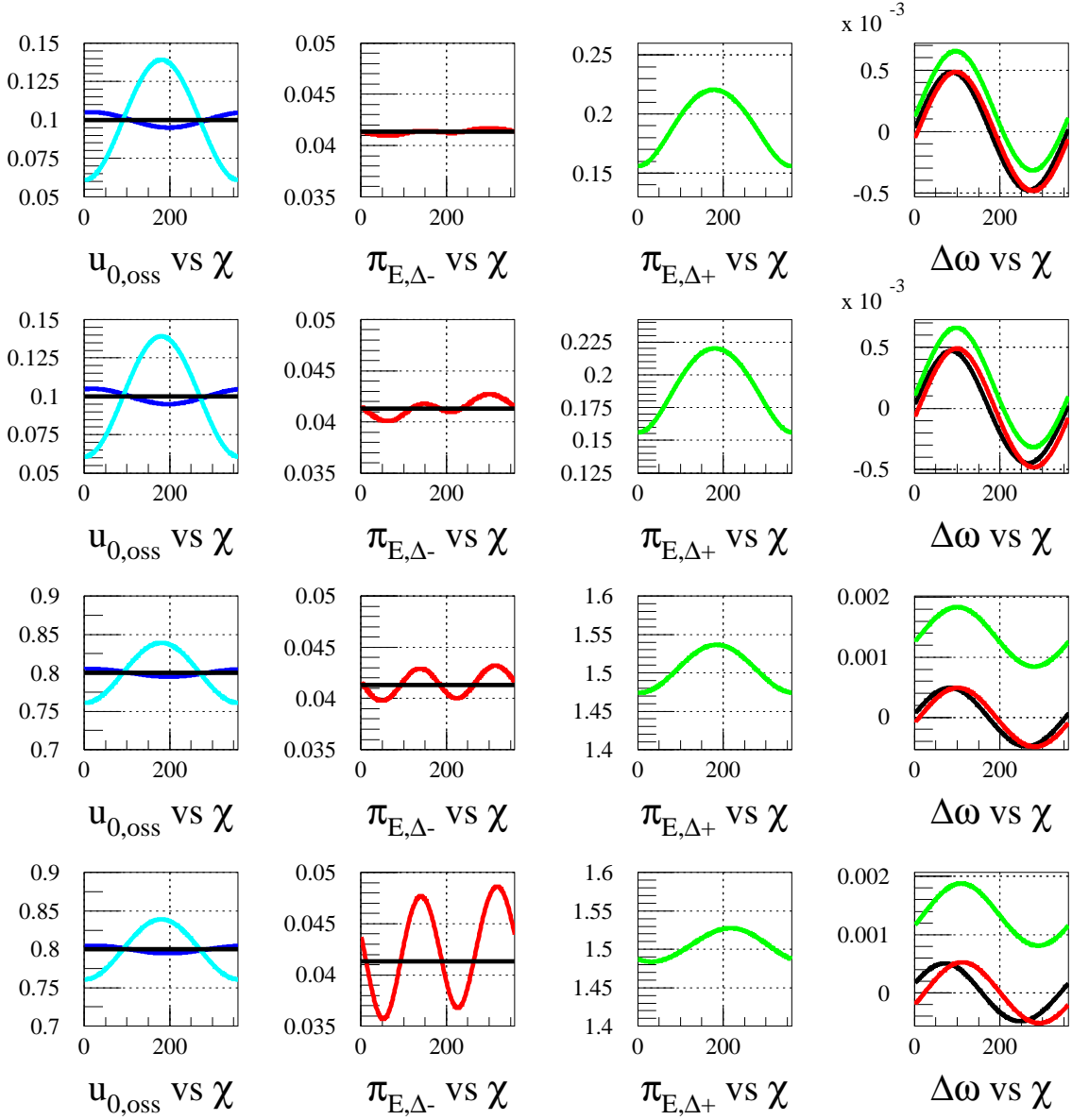}
\caption{For a Bulge lens, as a function of the direction
of the the lens trajectory, angle $\chi$ expressed in degree,
we show, from left to right, $u_{0,oss}$, 
as evaluated for observers in motion,
and the corresponding values
as evaluated assuming observers at rest for
$\pi_\mathrm{E,\Delta\pm}$ and $\Delta\omega$,
the true value in black, in red and green
those corresponding to $\pi_\mathrm{E,\Delta\pm}$.
The solid horizontal lines for $u_0$
and $\pi_\mathrm{E}$ indicate respectively
the impact parameter as would be seen 
from the Sun and the value of the
microlens parallax for the underlying microlensing event.
From top to bottom we show different configurations
increasing the value of the impact parameter
and the event duration. We refer to Section~\ref{sec:ana}
for full details.
}
\label{fig:p2bul}
\end{figure}

\begin{figure}
\epsscale{1.0}
\plotone{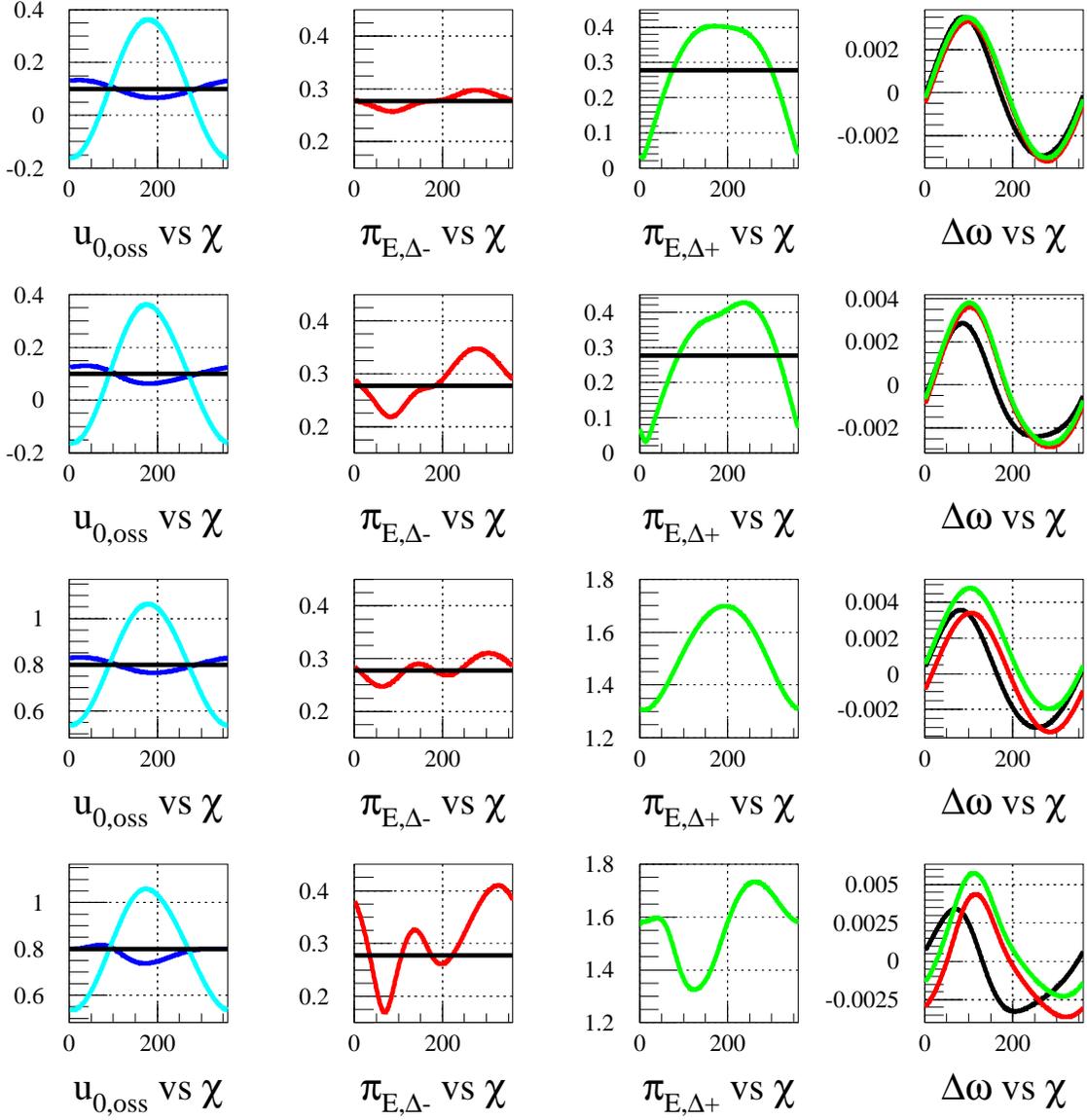}
\caption{The same as in Figure~\ref{fig:p2bul},
here for a lens in the Disc.  
We refer to Section~\ref{sec:ana}
for full details.
}
\label{fig:p2dis}
\end{figure}

We fix the line of sight towards the Bulge and $t_0$ as in 
Figures~\ref{fig:prest} and \ref{fig:prest2}.
Fixed the lens mass at $0.6~\mathrm{M}_\odot$ 
we consider two cases: a lens in the disc,
at $D_l = 2.0~\mathrm{kpc}$, and a lens 
in the Bulge, at $D_l = 7.5~\mathrm{kpc}$. This results,
always for $D_s=8.~\mathrm{kpc}$, in $\pi_\mathrm{E}=0.28$ and
0.04, respectively. For $v=300$ and $80~\mathrm{km/s}$,
the timescale is $16$ and $59~\mathrm{d}$ for the disc lens
and $8.7$ and $33~\mathrm{d}$ for the bulge lens.
In Figure~\ref{fig:p2bul} and \ref{fig:p2dis}
(bulge and disc lens, respectively)
we show for two values of the impact parameter, $u_0=0.1$ and $0.8$
(from top to bottom, for increasing impact parameter
and event duration),
as a function of the angle of the lens motion $\chi$,
the values for $u_{0,\mathrm{oss}}$, 
the horizontal solid line indicates $u_0$ for the Sun observer,  
those for $\pi_{\mathrm{E},\Delta\pm}$ as calculated
for observers at rest, where the solid horizontal line
indicates the true value for $\pi_\mathrm{E}$,
and $\Delta\omega$ corresponding to $\pi_{\mathrm{E},\Delta\pm}$.
Note that the two degenerate solutions
for the parallax amplitude, and correspondingly
the values for $\Delta\omega$,
are evaluated based on 
$\Delta_\pm \equiv \Delta u_{0,\pm} = u_{0,2} \pm u_{0,1}$,
namely taking into account the sign of $u_{0,\mathrm{oss}}$.
This way the $\Delta_-$ solution
has always  the same parity of the original configuration,
$\pi_{\mathrm{E},-}$ or $\pi_{\mathrm{E},+}$
(with the changes between the parity following the sign
of $u_{0,\mathrm{oss}}$).
Indeed we can see that the $\pi_{\mathrm{E},\Delta-}$
solution remains always close enough to the true value,
whereas $\pi_{\mathrm{E},\Delta+}$ can get quite different.
Interestingly, however, there are also
ranges of $\chi$ values for which
$\pi_{\mathrm{E},\Delta+}$ can get closer to the
true value than $\pi_{\mathrm{E},\Delta-}$,
so that in principle the analysis
based on the assumption of observers at rest
may lead to the correct value for $\pi_\mathrm{E}$
although with the wrong sign of the parity.
In general we see that a larger value
of $u_0$ or $t_\mathrm{E}$ makes both the difference between
the $\pi_{\mathrm{E},\Delta-}$ and $\pi_{\mathrm{E},\Delta+}$
values and the relative difference with respect
to the true value $\pi_\mathrm{E}$ larger,
as well as it makes larger the difference
between $\Delta\omega$ calculated for observers at rest
with respect to the true value.
However, whereas for the bulge lens
these differences do not ever become really significant, 
so that the lens motion may indeed be neglected
in the analysis, this is no longer true for the disc lens configuration.
It is apparent, therefore, that, besides
the effect of the direction of motion, larger values
of the microlens parallax (for large enough $u_0$),
therefore for decreasing values of the lens mass and
nearer lenses, tend to enhance the importance of the observers motion.
This effect is however balanced by the event duration
which on the other hand decreases both with the lens mass
and the lens distance.


\section{K2 and \emph{Spitzer} parallax: a 3 observers problem} 
\label{sec:k2}

K2 \citep{howell14}, the extension
of the \emph{Kepler} \citep{borucki10,koch10} mission, 
is expected to carry out in spring 2016
a three months microlensing monitoring towards the Galactic bulge
during its K2C9 campaign, the first space-based microlensing 
survey ever \citep{henderson15}.
The K2 survey mode of operation is opposed to that of the
\emph{Spitzer} observational
program which monitors microlensing
events in a follow-up mode \citep{yee15}.
This will allow K2 to address several
relevant scientific questions
related to the observation of a typology of microlensing
events, such as high magnification 
and/or short timescale ones,
which are likely to be missed by \emph{Spitzer}.

\emph{Kepler} is moving along an Earth-trailing orbit similar to that
of \emph{Spitzer} which therefore we approximate
in a similar way as in Section~\ref{sec:motion}
(the exact ephemeris can be found on the NASA-JPL Horizon system).
We fix the phase shifts at the fall equinox 2016,
$\mathrm{JD-2450000}=7654.1$,
for \emph{Spitzer} and K2 at 
$\Delta\lambda = -94^\circ.4,\,-52^\circ.7$, respectively.

K2C9 is going to last about three months,
starting April 2016, so that this
observational period will (partly) overlap
with that of the expected 2016 \emph{Spitzer}
follow-up microlensing project
expected to start in June 2016 (we recall that 
this campaign must obey the \emph{Spitzer}
visibility constraints towards the Bulge).
For the first time it is going to be possible to observe
simultaneously the same microlensing events
from ground and two satellites in orbit.
This is relevant to our analysis because, 
already within the framework of observers
at rest, the fourfold degeneracy
is broken by the introduction of a third observer,
an effect enhanced when correctly taking 
into account the motion of the observers.
It is however interesting to address 
this issue within the framework
of observers at rest to appreciate
to which extent the degeneracy gets actually broken.
From the inspection of Eq.~\ref{eq:tau} we see 
that the difference 
between the times at maximum amplification
is going to be different for K2 and \emph{Spitzer}, so that
the respective degenerate solutions for the lens trajectory
are going to be different.
Eq.~\ref{eq:gould94} then implies that
the difference in the impact parameters
will then determine two different  sets of $\pi_{\mathrm{E},\pm}$
solutions, giving therefore the chance,
when analysed together, to single out the correct one
(we recall that the degeneracy breaking
for K2 microlens parallaxes is the specific 
purpose of one of the accepted proposals \citep{spitzer16a}
for the forecoming 2016 \emph{Spitzer} campaign).

\begin{figure}
\epsscale{1.0}
\plotone{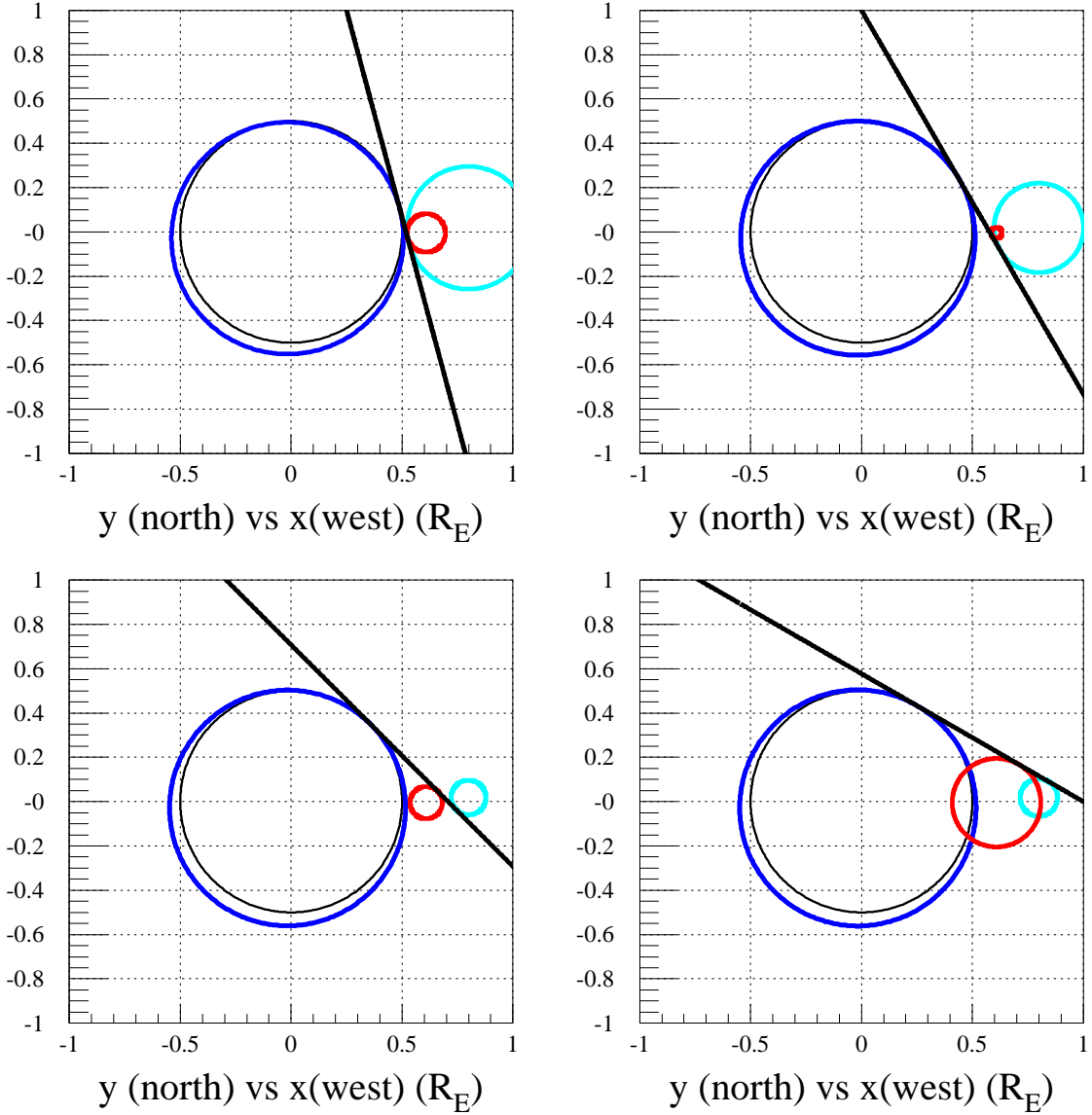}
\caption{Projected in the lens plane,
in an heliocentric reference frame
centered on the line of sight to the source,
the configurations, for four
different lens trajectories,
for a parallax microlensing event for three observers
at rest, Earth, \emph{Spitzer} and K2,
for a line of sight towards the Bulge
(see Section~\ref{sec:k2} for full details). 
The thin black circle is centered on the origin,
radius the impact parameter as would be seen from the Sun.
The thick circles are centered on the observer positions,
with radius the respective impact parameters,
dark, light blue and red for Earth, \emph{Spitzer} and K2.
The straight line represents the lens trajectory, simultaneously
tangent to all the circles.
The reference frame is as in Figure~\ref{fig:prest}.
}
\label{fig:k2a}
\end{figure}

This is exemplified in Figure~\ref{fig:k2a} 
where we show the configuration in the lens plane
for the three observers case 
for a specific microlensing event configuration
and 4 test values of the lens trajectory.
We show the circles of radius $u_{0,\mathrm{oss}}$
centered on the observers positions (here 
considered at rest, as evaluated at $t=t_0$) 
and the lens trajectory which is 
now simultaneously tangent to all three observer circles.
At glance it is clear that, 
when considering both the couples
of observers simultaneously, 
the degeneracy for the parallax vector
directions and amplitude is broken. 
For reference, we fix the line of sight
to that expected to be the center of K2C9 field,
$\mathrm{RA, DEC} = 270^\circ.354,\,-21^\circ.780$.
We fix $t_0=(\mathrm{JD}-2450000) = 7561$ (June 21, 2016),
$u_0=0.5$, $t_\mathrm{E}=24~\mathrm{d}$, $\pi_\mathrm{E}=0.8$
and test 4 values for the angle of the lens trajectory
(top to bottom, left to right)
$\chi=15^\circ,\,30^\circ,\,45^\circ$ and $60^\circ$.
The resulting $u_0$ from ground is always 
about $0.52-0.53$, for \emph{Spitzer}
0.28, 0.20, 0.078 and 0.084
and for K2 0.087, 0.025, 0.073 and 0.2;
$\Delta t_0$ is -4.0, -8.8,
-13., -16. days for \emph{Spitzer}
and -3.3, -7.0, -10 and -13 days for K2,
respectively. The degenerate directions, for each couple
Earth-\emph{Spitzer} and Earth-K2 are clearly
different. The resulting degenerate
values of the parallax amplitude are different
as well. In particular (for a true value
$\pi_\mathrm{E}=0.8$) for \emph{Spitzer} and K2 we evaluate
0.29 and 0.59 ($\chi=15^\circ$),
0.48 and 0.74 ($\chi=30^\circ$),
0.69 and 0.94 ($\chi=45^\circ$)
and 0.90 and 1.15 ($\chi=60^\circ$),
so that for this configuration the degenerate solution
shifts from $\pi_{\mathrm{E},-}$ 
to $\pi_{\mathrm{E},+}$ at $\chi=45^\circ$
for K2 and at $\chi=60^\circ$ for \emph{Spitzer}.
Clearly, the degree by which the degeneracy is broken
can be measured by how much
the degenerate solutions,
in term of direction or of the amplitude of $\vec\pi_\mathrm{E}$,
differ one from the other
when considering simultaneously the two couples of observers.
Hereafter we are going to focus on the amplitude
of the microlensing parallax.

For the analysis of Figure~\ref{fig:k2a} we have considered
the respective positions of the 
observers along their orbit, specifically
given their (fixed) phase shift.
This is indeed a relevant aspect
which leads us to discuss  
the seasonal effects for the measure
of the microlens parallax for two observers
lying both along the ecliptic plane.
Indeed, the line of sight towards the Bulge
is near the ecliptic plane therefore the projection
on the lens plane of the distance between
the two observers (which remains roughly constant
on the ecliptic plane along the few months of
a given observation campaign) is a strong function
of the period of the year. This is relevant because, 
from Eq.~\ref{eq:gould94},
we see that $\pi_\mathrm{E} \propto 1/D_\perp$.
All the microlensing parameters fixed, in particular
the parallax, whenever $D_\perp$ becomes very small,
namely when the two observers are roughly aligned
with the line of sight towards the Bulge,
the larger we can expect the degenerate
parallax solution to be. On the other hand,
$\pi_\mathrm{E} \propto \sqrt{1/M_l}\,\sqrt{(1-x)/x}$,
so that large values for $\pi_\mathrm{E}$ are expected
for very small lens mass or very nearby lenses.
Too extreme values (at least
for lenses in the stellar mass range), roughly $\pi_\mathrm{E}>2$,
are however by themselves extremely unlikely.

\begin{figure}
\epsscale{1.0}
\plotone{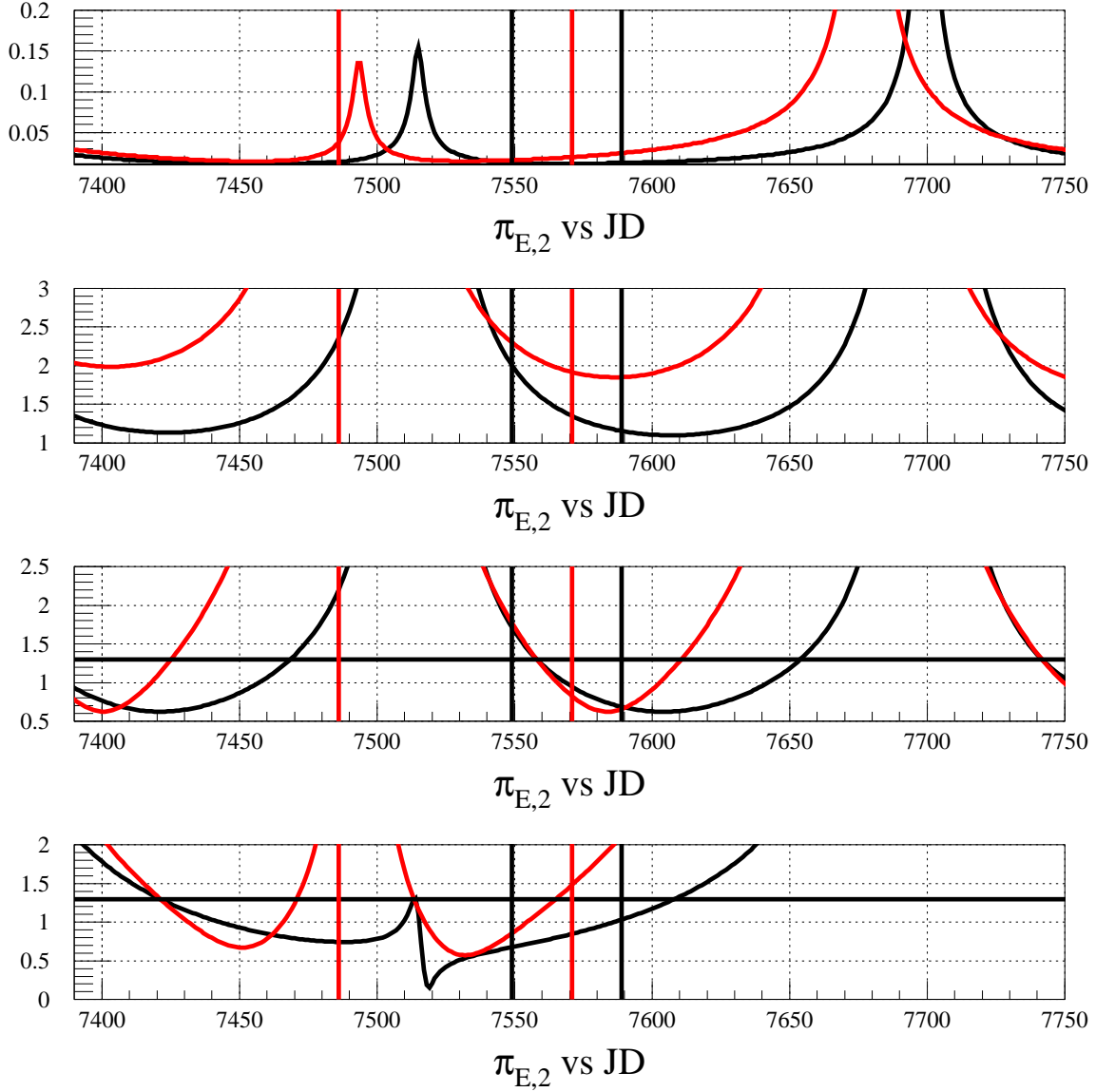}
\caption{The degenerate value of the
amplitude of the microlens parallax, $\pi_\mathrm{E,2}$,
for a given event configuration as a function of the
time at maximum magnification fixed at the date of the day
of the year 2016,
as measured from \emph{Spitzer} and K2,
black and red lines, observers at rest.
The horizontal lines indicate the true value
of the amplitude of $\pi_\mathrm{E}$.
Two top (bottom) panels
for $\pi_\mathrm{E}=0.01$ ($1.3$),
first and third (second and fourth) panels
from top for $u_0=0.01$ and $0.8$, respectively.
The vertical lines delimit the 2016
\emph{Spitzer} and K2 campaigns.
We refer to Section~\ref{sec:k2} for full details. 
}
\label{fig:k2b}
\end{figure}

In Figure~\ref{fig:k2b} we show the variation of the degenerate parallax
solution, $\pi_\mathrm{E,2}$ (which can be
$\pi_\mathrm{E,-}$ or $\pi_\mathrm{E,+}$)
as a function of the time
of the year, $\bar{t}$. Specifically, we fix
the  underlying microlensing event configurations
with in particular the time at maximum magnification,
as seen from the ideal observer on the Sun,
at $t_0=\bar{t}$. At the same time we fix the positions
along the orbit of the three observers,
Earth, \emph{Spitzer} and K2, at $t_0=\bar{t}$.
In particular, given the line of sight,
the direction of the lens motion and the
event timescale, we show the results
for different combinations of the impact
parameter and the microlens parallax.
For reference, the line of sight is fixed
as for the events shown in Figure~\ref{fig:k2a},
$t_\mathrm{E}=24~\mathrm{d}$ and $\chi=30^\circ$.
We then test two values for the microlens
parallax, $\pi_\mathrm{E}=0.01$ (two top panels) and 
$\pi_\mathrm{E}=1.3$ and two values for the 
impact parameter, $u_0=0.01$ (top to bottom, 
first and third panels =) and $u_0=0.8$.
At glance we can see that the difference in the orbital phase
shift between \emph{Spitzer} and K2 introduces
a shift in the peak for $\pi_\mathrm{E,2}$
and that, besides $\bar{t}$, also
the underlying microlensing configuration
plays a relevant role leading even to 
rather wild variations.
Focusing in particular on June 2016,
starting about JD-2450000=7550,
when the \emph{Spitzer} and K2 campaigns
will overlap, we see that, quite regardless
of the microlens parallax value,
small values for the impact parameter
tend to smooth over the difference
between the $\pi_\mathrm{E,2}$ values
as seen by \emph{Spitzer} and K2,
whereas larger values for $u_0$
lead quickly to a rather large offset
between the two $\pi_\mathrm{E,2}$ values
which can therefore resolve the parallax degeneracy.
Finally,  we note that for \emph{Spitzer},
within the boundaries of the observational window,
$\pi_\mathrm{E,2}$ tend to remain always roughly constant,
or in any case bounded to smaller value,
which is not the case, however, for K2.


\section{Conclusion}

In this work we have revisited the
analysis of the microlensing parallax
for the case of the simultaneous observation
of the same microlensing event by two, and three, observers
\citep{refsdal66,gould94b}
from within an heliocentric frame.
The main purpose of this analysis
is the understanding of the fourfold microlensing
parallax degeneracy and of how it is broken.
First we have discussed the case for observers at rest
and went through the geometrical meaning
of the microlensing parallax degeneracy,
in particular we have explicitly written down
an expression for the degenerate directions
of the lens trajectory as a function 
of the microlensing parallax, $\pi_\mathrm{E}$,
and $\tau=\Delta t_0/t_\mathrm{E}$ only.
Second, we have considered the case for observers
in motion and we have
written down an extension
to this case of the \cite{gould94b}
relationship between the microlensing
parallax and the light curve observables.
We have discussed
how the geometry of the microlensing parallax configuration
is now determined by all the parameters
of the underlying microlensing event,
in particular the duration,
which is the underlying reason 
for the breaking of the degeneracy in this case.
Through all the analysis, the choice of an
heliocentric reference frame allowed us
to get a clear geometrical and analytical insight.

As test case we have considered simultaneous
observations from ground and \emph{Spitzer},
relevant in consideration of the ongoing follow-up
observational campaign towards the Galactic Bulge
\citep{spitzer14,spitzer15,spitzer16a,spitzer16b}.
The analysis, through a series of test cases, 
hints that the motion of the observers can be
expected to be relevant especially for disc lenses
with large enough impact parameters and long enough timescale.
These are the cases, therefore, for which one may expect
to be able to break more easily the degeneracy
also from an observational point of view.

Finally, we have discussed the case
for three simultaneous observers,
relevant for the foreseen K2
microlensing survey expected for 2016,
which will also partly overlap with the \emph{Spitzer} season.
The microlensing parallax degeneracy is there broken
already from the standpoint of an analysis
based on the assumption of observers at rest.
Through a series of test cases we have shown
how this can be actually effective.

In this work we  explicitly have not addressed
the issue of the actual determination 
of the microlensing parameters out of observed light curves. 
We  recall that as a standard,
for instance for the analysis of the
\emph{Spitzer} light curves \citep{novati15}
the scheme developed by \cite{gould04}
from within a geocentric frame is used.
A drawback of this approach is that 
in the end one has to come
back to the heliocentric frame for the determination
of some parameters, notably the event duration.
A main advantage, however,
is that the observed underlying event parameters,
time of maximum magnification,
impact parameter and duration,
are similar to those which can be estimated
from ground in absence of the parallax effect.
In principle, though, this is exactly where
the  simultaneous observation from a space observer may help. 
With reference to the \emph{Spitzer} campaign,
however, we recall that,
even put aside the problems related
to the determination of the source flux
and the different blend fractions,
the limited baseline in most cases does not
allow to fully independently characterize
the microlensing light curve.
This represents a major problem for
the practical application of the analysis
presented in this work,
and this holds in particular
for the case of the three observatories.
Still, it is going to be interesting 
to analyse the K2 data which, thanks
to the survey mode and longer baseline available, 
may be expected to suffer less from this limitation.

\acknowledgments
We thank A.~Gould for valuable discussions.  
SCN acknowledges support by JPL grant 1500811.
GS thanks NExScI for hospitality at Caltech 
during part of this work.

\bibliographystyle{apj}
\bibliography{biblio}

\end{document}